\documentclass[twocolumn]{aastex631}

\usepackage{CJKutf8}

\newcommand{\cntext}[1]{\begin{CJK}{UTF8}{gbsn}#1\end{CJK}\kern-1ex}
\shorttitle{Multiple electron acceleration}
\shortauthors{Battaglia et al.}

\begin{document}

\correspondingauthor{Marina Battaglia}
\email{marina.battaglia@fhnw.ch}

\title{Multiple electron acceleration instances during a series of solar microflares observed simultaneously at X-rays and microwaves}
\author[0000-0003-1438-9099]{Marina Battaglia}
\affiliation{University of Applied Sciences and Arts Northwestern Switzerland, 5210 Windisch, Switzerland}

\author[0000-0003-0485-7098]{Rohit Sharma}
\affiliation{University of Applied Sciences and Arts Northwestern Switzerland, 5210 Windisch, Switzerland}

\author[0000-0002-5431-545X]{Yingjie Luo (\cntext{骆英杰})}
\affiliation{Center for Solar-Terrestrial Research, New Jersey Institute of Technology, 323 M L King Jr Blvd, Newark, NJ 07102-1982, USA}
\author[0000-0002-0660-3350]{Bin Chen (\cntext{陈彬})}
\affiliation{Center for Solar-Terrestrial Research, New Jersey Institute of Technology, 323 M L King Jr Blvd, Newark, NJ 07102-1982, USA}

\author[0000-0003-2872-2614]{Sijie Yu (\cntext{余思捷})} 
\affiliation{Center for Solar-Terrestrial Research, New Jersey Institute of Technology, 323 M L King Jr Blvd, Newark, NJ 07102-1982, USA}

\author[0000-0002-2002-9180]{S\"am Krucker}
\affiliation{University of Applied Sciences and Arts Northwestern Switzerland, 5210 Windisch, Switzerland}
\affiliation{Space Sciences Laboratory, University of California, 7 Gauss Way, 94720 Berkeley, USA}

\begin{abstract}
Even small solar flares can display a surprising level of complexity regarding their morphology and temporal evolution. Many of their properties, such as energy release and electron acceleration can be studied using highly complementary observations at X-ray and radio wavelengths. We present X-ray observations from the Reuven Ramaty High Energy Solar Spectroscopic Imager (RHESSI) and radio observations from the Karl G. Jansky Very Large Array (VLA) of a series of GOES A3.4 to B1.6 class flares observed on 2013 April 23. The flares, as seen in X-ray and extreme ultraviolet (EUV), originated from multiple locations within active region NOAA 11726. A veritable zoo of different radio emissions between 1 GHz and 2 GHz was observed co-temporally with the X-ray flares. In addition to broad-band continuum emission, broad-band short-lived bursts and narrow-band spikes, indicative of accelerated electrons, were observed. However, these sources were located up to 150 arcsec away from the flaring X-ray sources but only some of these emissions could be explained as signatures of electrons that were accelerated near the main flare site. For other sources, no obvious magnetic connection to the main flare site could be found. These emissions likely originate from secondary acceleration sites triggered by the flare, but may be due to reconnection and acceleration completely unrelated to the co-temporally observed flare. Thanks to the extremely high sensitivity of the VLA, not achieved with current X-ray instrumentation, it is shown that particle acceleration happens frequently and at multiple locations within a flaring active region.   
\end{abstract}


\section{Introduction}
Solar flares are magnetic energy release events in the solar atmosphere that efficiently accelerate particles to relativistic speeds. The most direct signatures of flare accelerated electrons are found at X-ray and radio wavelengths. Observations at these wavelengths are highly complementary \citep[e.g.][]{2011SSRv..159..225W} and allow us to diagnose the locations and mechanisms of particle acceleration in solar flares and to study how particles are transported close to the Sun and away from the Sun. \\
\begin{figure*}[!]
\centering
\includegraphics[width=0.9\linewidth]{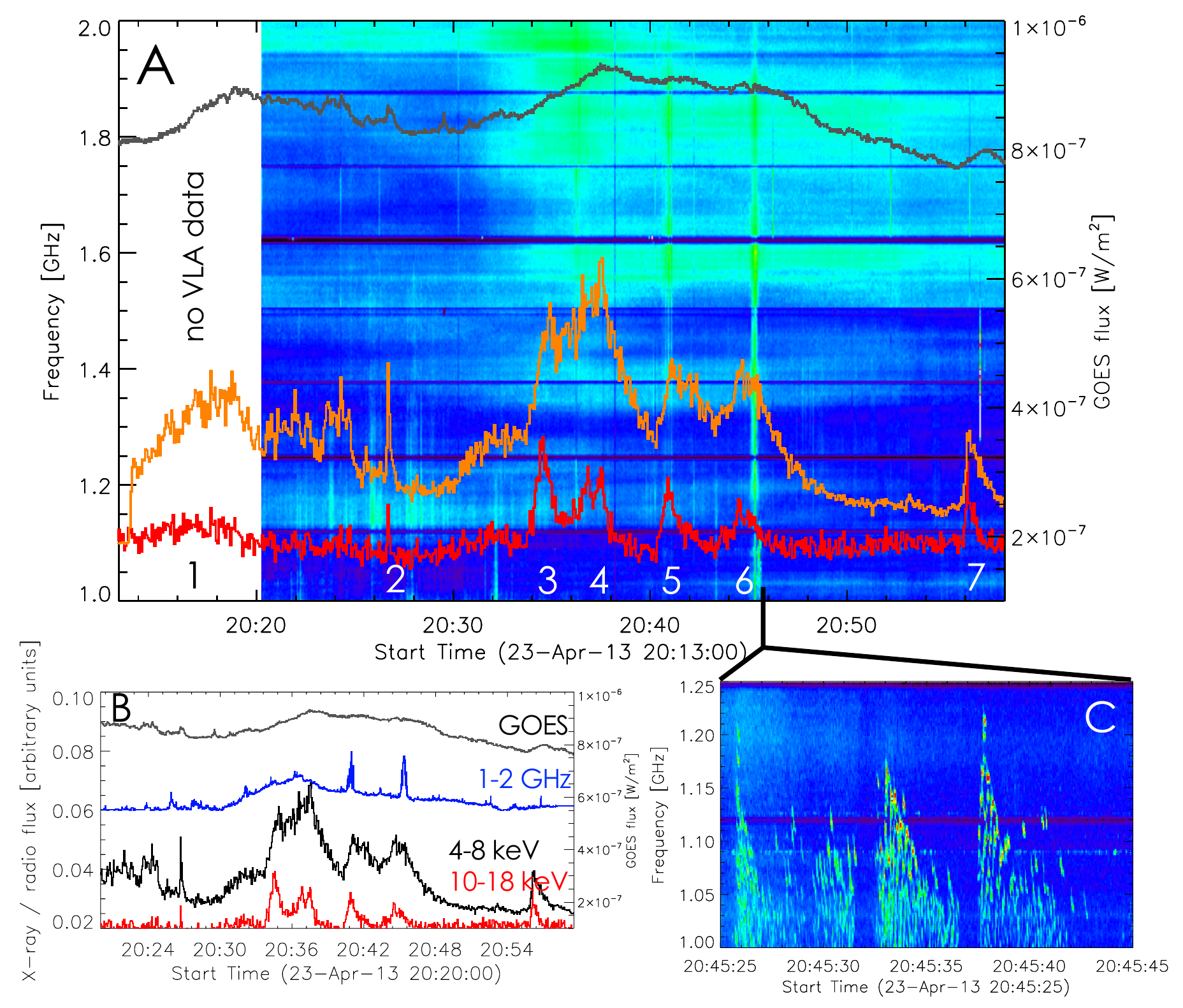}
\caption{Panel A: VLA median cross-power dynamic spectrum between 1 GHz and 2 GHz. The orange and red lines are the RHESSI 4$-$8~keV and 10$-$18~keV lightcurves, respectively (arbitrary scaling). Flares were identified as individual peaks in the 10$-$18~keV lightcurve and labelled 1 to 7. The GOES lightcurve at 1$-$8~\AA\ is given in dark grey (axis on the righthand side). Panel B: X-ray lightcurves as in Panel A. Additionally, the frequency integrated radio lightcurve is given as blue line. Panel C: Inset between 20:45:25 UT and 20:45:45 UT, showing narrow-band spikes in the decay phase of F6.}
\label{fig:lc}
\end{figure*}
Gyrosynchrotron emission at frequencies above $\sim$ 2 GHz and X-ray emission at a few tens of keV originating from high energetic, nonthermal electrons are often found to be tightly correlated \citep[e.g.][]{2002ApJ...580L.185M,2010ApJ...714.1108K}. In a statistical study on the timings of X-ray and radio emission at 17 GHz and 34 GHz, \citet{2020ApJ...894..158K} found that this may in fact hold for the majority of larger flares. In single-event analysis, such common origin of the emissions leads to a more complete scenario of electron acceleration than observations at one wavelength alone \citep[e.g.][]{2018ApJ...863...83G}. \citet{2021ApJ...908L..55C} performed an in-depth study of a source above the top of magnetic loops (above-the-loop-top source). Such sources have been interpreted as the (primary) acceleration site \citep[e.g.][]{1994Natur.371..495M,2010ApJ...714.1108K,2012ApJ...748...33C,2014ApJ...780..107K,2019ApJ...872..204B}. Combining spatially resolved X-ray and radio spectra, \citet{2021ApJ...908L..55C} inferred the energetic electron spectrum at the acceleration site over a much wider energy range than would be possible from observations at either wavelength alone. \\
However, very often, radio emission is not observed from the same location as X-rays. This is typically the case for different types of coherent emission observed below $\sim 2$~GHz. In many instances, such emission is due to beams of accelerated electrons whose X-ray signatures are too faint to be detected with current instrumentation \citep[e.g.][]{2009ApJ...696..941S,2018ApJ...867...84G,2018ApJ...866...62C}, that are propagating away from the Sun on open field lines. In many of these cases it is found that the electrons were accelerated at the same site as electrons that become trapped in closed magnetic loops. However, a number of authors found such emissions to originate from a different, secondary acceleration site \citep[e.g.][]{2009A&A...499L..33B, 2011SoPh..273..363B, 2016ApJ...833...87C, 2015Sci...350.1238C, 2019ApJ...884...63C}. 
Large solar flares usually have a long duration (up to several hours) and often display a complex morphology with multiple loop systems, hence it is not surprising that multiple acceleration sites are found. However, even smaller flares, so-called microflares \citep{2011SSRv..159..263H} can display a remarkable complexity as demonstrated by e.g. \citet[][]{2020ApJ...904...94S}, \citet{2020ApJ...891L..34G}, \citet{2021ApJ...908...29D}, and \citet{2021ApJ...913...15V}.
Understanding this complexity and different features observed at both, radio and X-ray wavelengths, is the key to understanding where and how electron acceleration takes place and how electrons are transported. 

Here, we present a series of microflares, ranging in GOES class from B8 to B9.3 (A3.4 to B1.6 background-subtracted), that were observed simultaneously at X-rays by the Reuven Ramaty High Energy Solar Spectroscopy Imager \citep[RHESSI,][]{2002SoPh..210....3L}, at radio frequencies between 1 GHz and 2 GHz by the Karl G. Jansky Very Large Array \citep[VLA,][]{2011ApJ...739L...1P}, and at extreme ultraviolet (EUV) wavelengths by the Atmospheric Imaging Assembly \citep[AIA,][]{2012SoPh..275...17L} on board the Solar Dynamics Observatory \citep[SDO,][]{2012SoPh..275....3P}. The flares happened on 2013 April 23 over a duration of fifty minutes. They all originated from active region NOAA 11726 (N13W49) but from different locations within it. During the observations, a veritable zoo of different radio emissions was observed, most, but not all, temporally associated with X-ray flares. However, there was no spatial association between the radio emissions and the X-ray flare locations, suggesting multiple acceleration instances and sites before, during and after the main flares, and energetic particles that gained access to large or open field lines far away from the main acceleration site. This work is intended as an overview of all events, representing the "bigger picture". Individual events will be analysed in-depth in a follow-up publication. 

In Section~\ref{s:overview} an overview of the active region, the X-ray flares and radio emissions is given, followed by a detailed description of imaging and spectral analysis in both X-rays and radio wavelengths (Section \ref{s:analysis}). In Section~\ref{s:results}, a detailed investigation of the temporal and spatial association between radio emissions and X-ray emission in the different flares is made. The findings are discussed and interpreted in Section~\ref{s:discuss}. Conclusions are drawn in Section~\ref{s:conc}
\begin{table*}
\begin{tabular}{ccccc}
\hline
\hline 
Flare & X-ray & X-ray & Dominant Radio feature & Radio image \\
No. & peak-time (UT) & integration time (UT) & in dynamic spectra & integration time \\ 
\hline
1 & 20:17:42 & 20:15:00 - 20:19:30 & no VLA data & $-$  \\
2 & 20:26:10 & 20:27:10 - 20:29:10 & none & 20:26:41.5-20:26:42.5\\
3 & 20:34:25& 20:33:30 - 20:35:36 & broad-band continuum & 20:34:30-20:34:31\\
4 &20:36:47 & 20:35:40 - 20:38:36 & broad-band continuum &  20:37:31-20:37:32\\
5 & 20:40:51& 20:40:00 - 20:42:36 & short-lived burst & 20:40:54 - 20:40:55 \\
6& 20:44:31& 20:43:50 - 20:46:02 & none & 20:44:30-20:44:31  \\
6b & 20:44:31& & short-lived burst during decay phase & 20:45:21-20:45:22\\
6s & 20:44:31& & narrow-band spikes during decay phase & 20:45:25.90 - 20:45:25.95 \\
7& 20:56:04& 20:55:20 - 20:57:00 & none & 20:56:05.5 - 20:56:06.5 \\
\hline
\end{tabular}
\caption{X-ray peak-times in 10$-$18 keV lightcurve, X-ray image intergration time, radio feature identified in dynamic spectrum and radio image integration times.}
\label{t1}
\end{table*}
\section{Event overview} \label{s:overview}
The flares were observed on April 23 2013 between 20:10 UT and 21:00 UT. Figure~\ref{fig:lc} shows the median radio cross-power dynamic spectrum between 1 and 2 GHz along with the GOES 1$-$8 \AA\ lightcurve and RHESSI X-ray lightcurves at two energy bands (4$-$8~keV and 10$ -1 8$~keV). Seven individual flares were identified based on X-ray emission in the RHESSI 10$ - $18~keV lightcurves, the highest photon energies that RHESSI detected from these flares. The individual flares were numbered 1 to 7, as indicated in the Figure. We will refer to them as F1 through F7. The corresponding radio sources will be denoted F1$_R$ through F7$_R$. For all flares, except F3 and F4, the intensity at 10 $-$ 18 keV returned to background level between each flare. Since X-ray images of F4 showed an additional source, not visible during F3, these two peaks were treated as individual flares rather than one flare with two peaks. The 10-18~keV emission of all flares except F1 is quite impulsive, suggesting potential nonthermal emission while the lightcurve of F1 is more gradual in both energy ranges, hinting at purely thermal emission. A definite distinction is only possible through spectroscopy (see Section \ref{xrayspec}).

During the whole duration of the observations, various types of radio emissions were visible in the dynamic spectrum:  broad-band continuum emission above 1.5 GHz, most prominently during F3 to F7, broad-band short-lived bursts during F5 and F6, and narrow-band spikes in the decay phase of F6. The latter two will be referred to as F6b$_R$ and F6s$_R$. Even though these emissions were temporally associated with the X-ray flares, there was no clear time correlation. Table~\ref{t1} gives as summary of the observed features, including the X-ray peak-times, the time integrations of the RHESSI and VLA images (see Section~\ref{s:analysis}) and a short description of the observed features in the radio dynamic spectrum. 

\section{Data analysis}\label{s:analysis}
X-ray images and spectra were generated at the high-energy X-ray peak-time of each of the seven flares. Radio images and spectra were generated at the same times as the X-ray data products. Additionally, images and spectra of the burst and the spikes in the decay phase of F6 were generated, as described in detail below. 
\subsection{X-ray imaging and spectroscopy}\label{xrayspec} \label{subs:xrayspec}
RHESSI images were generated using the CLEAN algorithm \citep{Hog74,2002SoPh..210...61H} for two energy bands: $4-8$~keV (thermal emission) and $10-18$~keV (potentially nonthermal emission). For the image reconstruction, detectors 5, 6, 7, and 8 were chosen to limit noise contributed by detectors with finer grids either because there was no flux modulation or because of reduced detector sensitivity during that time of the mission. The image integration time was chosen to include the whole high-energy peak but minimum 60~s to achieve good count statistics, especially in the higher energy band. The image integration times are listed in Table~\ref{t1}. X-ray spectra were generated using the same integration times as for the images. For spectral fitting, detector 6 was used as, at that point in the RHESSI mission, this detector had the highest sensitivity with relatively low background. All flares were fitted with a single thermal component at the lowest energies (compare Table~\ref{t2}). However, spectral fitting of the higher energies proved difficult. After careful background subtraction, only a few counts were observed at energies between 10$-$18 keV, most prominently in F4 and F7. Since detector pileup could be excluded as origin of this emission, we interpreted it as weak signatures of accelerated, nonthermal electrons and took the following approach to obtain an upper limit on the nonthermal electron spectrum. The single thermal component fitted previously was fixed and an additional nonthermal thick-target component was fitted up to $\sim$18~keV, depending on the flare. Then both, the thermal and nonthermal component were fitted simultaneously. Even though a nonthermal component could be fitted in this manner, the parameters were not well constrained with uncertainties that were larger than the fit parameters in many cases. 
An example spectrum is given in Figure~\ref{fig:bgburst}. The nonthermal spectrum shown in the figure corresponds to a total electron flux of $F_e=1.0\times 10^{34}$s$^{-1}$, electron spectral index $\delta=4.9$ and low-energy cutoff $E_{low}=6.5$ and has to be seen as an upper limit. An additional difficulty was posed by  F1 and F4 for which two X-ray sources could be imaged simultaneously. Hence, these flares actually constituted of two separate events happening simultaneously and the full Sun spectra are superpositions of both. For larger flares, individual spectra of flaring sources can be isolated through imaging spectroscopy \citep[e.g.][]{2003ApJ...595L.107E,2006A&A...456..751B,2013A&A...551A.135S,2014ApJ...780..107K}. However, the sources here were too faint to be imaged at more than two energy bands. Note that a fitting model consisting of two purely thermal components with different temperatures would be another possibility. Such a model resulted in even larger uncertainties of the fit parameters and was discarded. 
\begin{table}
\begin{tabular}{cccc}
\hline
\hline 
Flare & Emission measure & Temperature \\ 
No. & $(10^{46}\rm{cm^{-3}})$ & (MK) \\
\hline
1 & $4.1\pm   0.3$  &$ 10.7 \pm 0.1$  \\
2&   $4.5 \pm  0.5$  & $10.0  \pm 0.2$  \\
 3& $3.5  \pm 0.3$ & $11.4  \pm 0.1$  \\
  4& $ 1.3 \pm0.9$& $14.3  \pm 1.7 $  \\
 5& $ 1.3 \pm0.9$& $14.3  \pm 1.7 $  \\
    6& $  1.7 \pm  0.1$ & $12.6  \pm 0.2$ \\
 7&  $ 1.2 \pm1.8$ & $11.5 \pm  0.7$  \\
\hline
\end{tabular}
\caption{Temperature and emission measure from X-ray spectral fitting}
\label{t2}
\end{table}
\subsection{Radio imaging and spectroscopy}
The VLA observations (under the observing program VLA/13A-384) were carried out using two subarrays. One subarray observed in the 1--2 GHz ($\lambda = 15$--30 cm) L band with 13 antennas, and another in the 2--4 GHz ($\lambda = 7.5$--15 cm) S band with 13 antennas. In the present study, we focus on the 1--2 GHz L band data where the radio bursts are present. The L band data had a spectral resolution of 2 MHz and temporal resolution of 50 ms in both the right-hand- and left-hand-circular polarization (RCP and LCP). The spectral range of the observations was divided uniformly into 8 spectral windows. Each spectral window had 64 2-MHz-wide frequency channels.  The L-band subarray was in D-configuration with a total of 13 antennas used for the observations. The longest baseline was 1031 m, corresponding to an angular resolution of $70'' \times 40''$ at $1.5$ GHz at the time of the observations. 

\begin{figure*}
\centering
\includegraphics[width=0.9\linewidth]{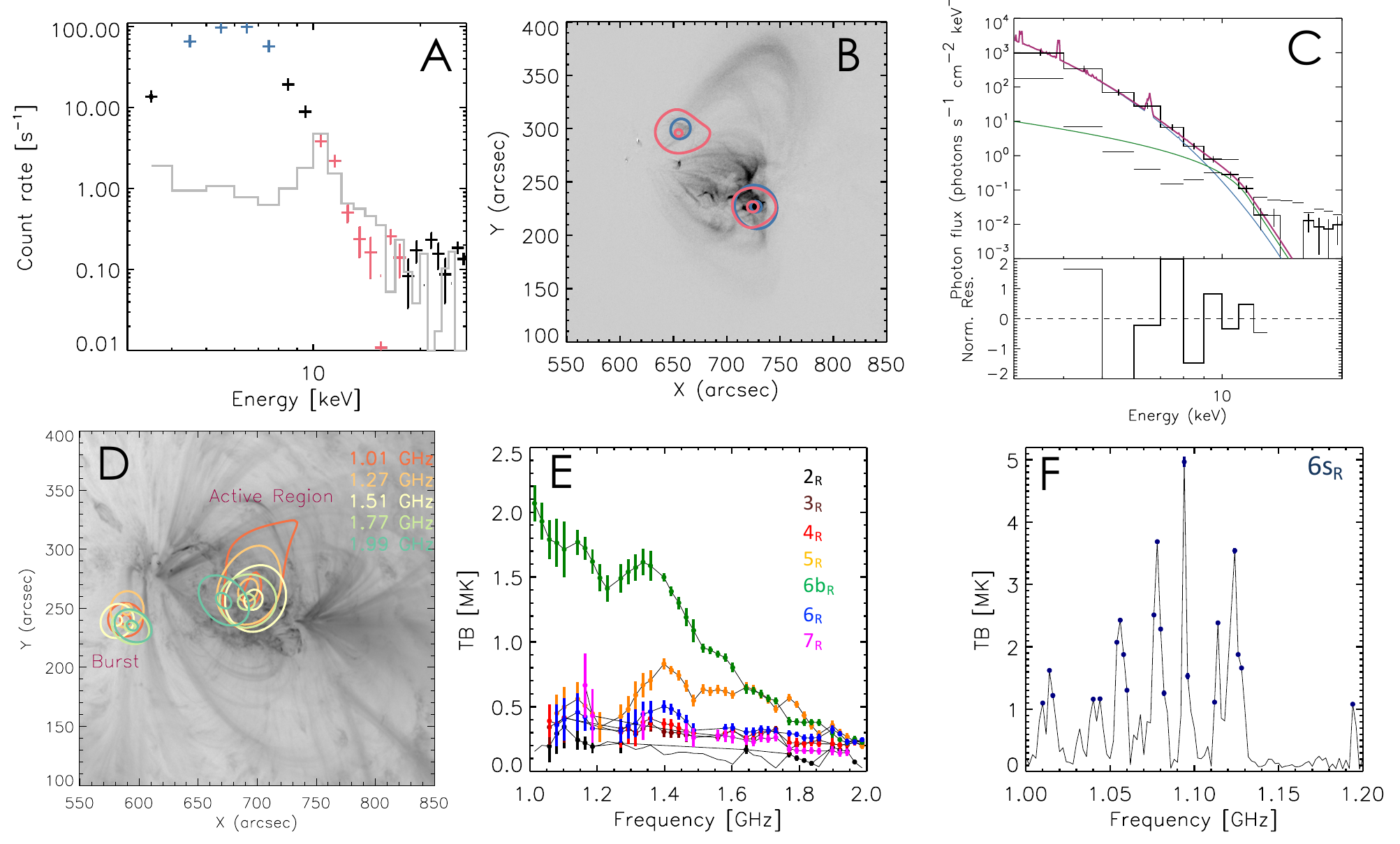}
\caption{Panels A and B: Example of RHESSI count spectrum and corresponding image (F4). Panel A shows the RHESSI count spectrum where the range $4-8$~keV is colored blue and the range $10-18$~keV is colored pink. The grey line gives the background level. The image in panel B is an AIA \ion{Fe}{18} image with contours from the RHESSI images at $4-8$~keV and $10-18$~keV overlaid in the respective colors. In this case, the spectrum and images suggest some flaring emission, of the same order as the background level, up to 15~keV. Panel C gives the photon spectrum fitted with a thermal component (blue) plus a nonthermal thick target component (green). The red curves represents the combined model. Panel D: Comparison of the radio source locations at selected frequencies (colors) between active region thermal emission and the radio burst during F6. Contours are at 90\% and 99\% levels of maximum emission in each image. Panel E: Radio brightness temperature spectra, color coded according to legend, where 6b$_R$ denotes the radio burst during F6. Colored dots indicate the frequencies for which a source position was determined. Panel F: Radio brightness temperature spectra of the short-lived spikes (F6s$_R$) during F6.}
\label{fig:bgburst}
\end{figure*}

CLEAN images in both RCP and LCP were made during the X-ray peak-time of each flare and, additionally, for the burst and the spikes in the decay phase of F6. A time-integration of one second and frequency integration of 20 MHz were used for all images except the spikes. The latter were only detected at the lowest frequencies, extremely short-lived and much brighter than the other sources. Hence, they were imaged at full time resolution of 50~ms and full frequency resolution of 2~MHz. A circular beam with FWHM of 70 arcsec at 1 GHz, linearly decreasing to 40 arcsec at 2 GHz, was used for image convolution. The radio emission observed in the dynamic spectrum was dominated by background emission of the active region that acted as a strong background source. Hence, background subtraction was performed in the visibility domain to isolate the flaring sources. As background time, the time interval between 20:08:00 and 20:08:01 UT was chosen (20:45:24.5-20:45:25.0 UT for F6s$_R$). Figure \ref{fig:bgburst} shows images of the background source and the burst in LCP during F6 at selected frequencies for comparison. 
The centroid location of the burst sources was determined as a function of frequency by fitting a 2D Gaussian at a flux level of $>90\%$ of the maximum emission in each image.
Radio spectra were computed from the brightness temperature maps by selecting the maximum brightness temperature of the source in images for which a source location could be fitted. At some frequencies the images were dominated by noise and no location could be fitted. In that case, no data point is shown in Figure~\ref{fig:bgburst}. The uncertainties of the brightness temperature spectra are given as the root-mean-square noise of the images, excluding the flaring source.  

Imaging and extraction of spectra was done for both polarizations. However, only the spikes during F6 were highly polarized ($<-$0.8), where the degree of circular polarization was defined as (RCP-LCP)/(RCP+LCP). F6b had a polarization degree between $-$0.7 and 0.2, depending on frequency. All other sources were polarized to less than 30\%. For further analysis and interpretation, we used the LCP data. 
\subsection{EUV and magnetic field imaging}
AIA images and magnetic field maps from the Helioseismic and Magnetic Imager \citep[HMI,][]{2012SoPh..275..207S} on SDO were processed using the standard data processing pipeline for these instruments. Using the method developed by \citet{2013A&A...558A..73D}, we also constructed AIA maps of the \ion{Fe}{18} line, which has a peak formation temperature of 8~MK \citep{2015A&A...582A..56D}, typical for smaller flares.  
\section{Results}\label{s:results}
First, we investigate the X-ray location of each flare relative to the other flares and to flaring emission seen at EUV wavelengths and an HMI magnetogram, followed by presentation of the radio source locations relative to each-other and to the X-ray sources and EUV emission features.
\subsection{X-ray source locations}
\begin{figure*}
\centering
\includegraphics[width=0.8\linewidth]{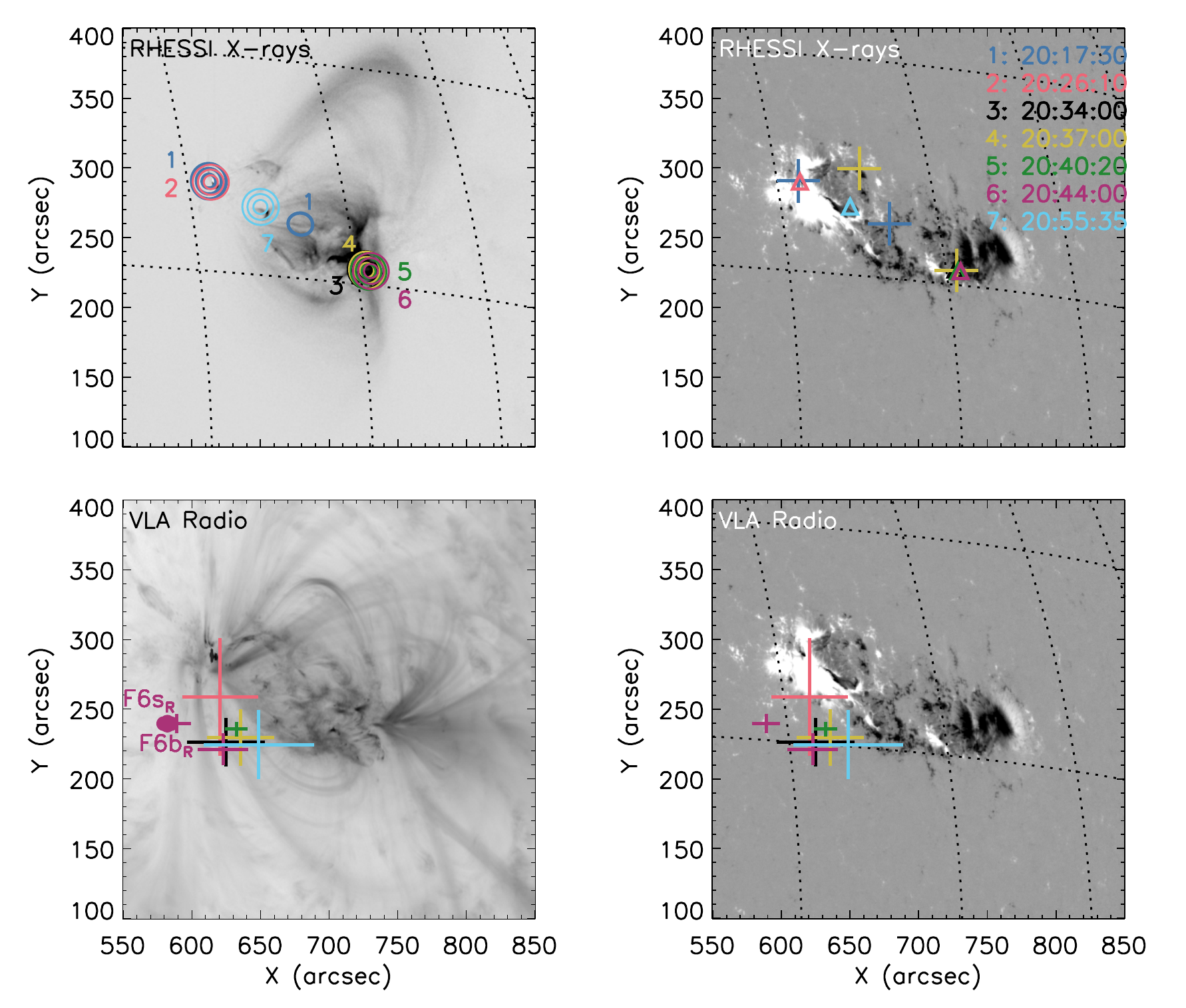}
\caption{Top row: Locations of X-ray sources at 4$-$8 keV relative to EUV emission in the \ion{Fe}{18} line (left) and an HMI magnetogram (right).  On the AIA F\ion{Fe}{18}-image the X-ray sources are indicated by the 50\%, 70\%, and 90\% contours of RHESSI CLEAN. Numbers next to the sources give the flare number according to the legend on the right. On the right, the location of maximum emission in RHESSI images is indicated by crosses. Here, the location of weaker X-ray sources during F1 and F4 are also indicated near coordinates [650,300] and [680,260], respectively. Different symbols are used for better visibility with arbitrary symbol size. The starting times of the RHESSI images are given to the right in the respective colors along with the flare number. Bottom row: AIA 171 \AA\ image and HMI image overlaid with the frequency-averaged location of the radio sources associated with each flare. The size of the cross gives the standard deviation of the frequency-dependent source locations. For better visibility, the location and standard deviation of the spike locations is indicated with an ellipse. The locations of F6b$_R$ and F6s$_R$ are marked, respectively.}
\label{fig:overviewimg}
\end{figure*}
Figure~\ref{fig:overviewimg} summarizes the locations of the X-ray flares relative to features in AIA 171 \AA\ and \ion{Fe}{18} images and an HMI magnetogram. The AIA images were composed by averaging images at 20:27:01 UT, 20:45:01 UT, and 20:54:59 UT. The AIA 171 \AA\ wavelength band is most sensitive to cooler plasma around 1 MK while the \ion{Fe}{18} images show plasma at 8 MK.  The locations of the X-ray sources at 4$-8$ keV are given as contour levels relative to the maximum in each image on the AIA image and as symbols on the HMI image. The magnetogram shows a complex magnetic topology with large positive and negative polarity patches at either side of the active region, interspersed with smaller patches of either polarity. EUV images show multiple loops at different temperatures. A large loop system, best visible in \ion{Fe}{18} images, seems to connect the large positive and negative magnetic field patches. Smaller loops are visible between the large loops. X-ray and radio images show sources from multiple locations within the active region over the fifty minutes of observations. F1 and F2 originated from the base of a larger loop system, visible at 171 \AA\ and where associated with an EUV jet. There was a second, fainter source imaged during F1 that was associated with a different loop system seen in the \ion{Fe}{18} images. F3 to F6 originated from a small loop system, clearly visible in the \ion{Fe}{18} images but not at the same location as F1 and F2. An additional, fainter source was observed during F4 that had no such obvious EUV counterpart. F7 originated from yet another location, close to the footpoint of an EUV loop.

\subsection{Radio source locations}
Figure~\ref{fig:overviewimg} gives the frequency-averaged centroid locations of the radio sources during each X-ray flare. The error bars give the $1\sigma$ standard deviation, representing the frequency-dependent scatter of the sources. All radio sources were located Eastward of the hot EUV loops and the locations with strongest magnetic field. None of them were co-spatial with any of the X-ray sources nor with the EUV loops.

\begin{figure}
\centering
\includegraphics[width=\linewidth]{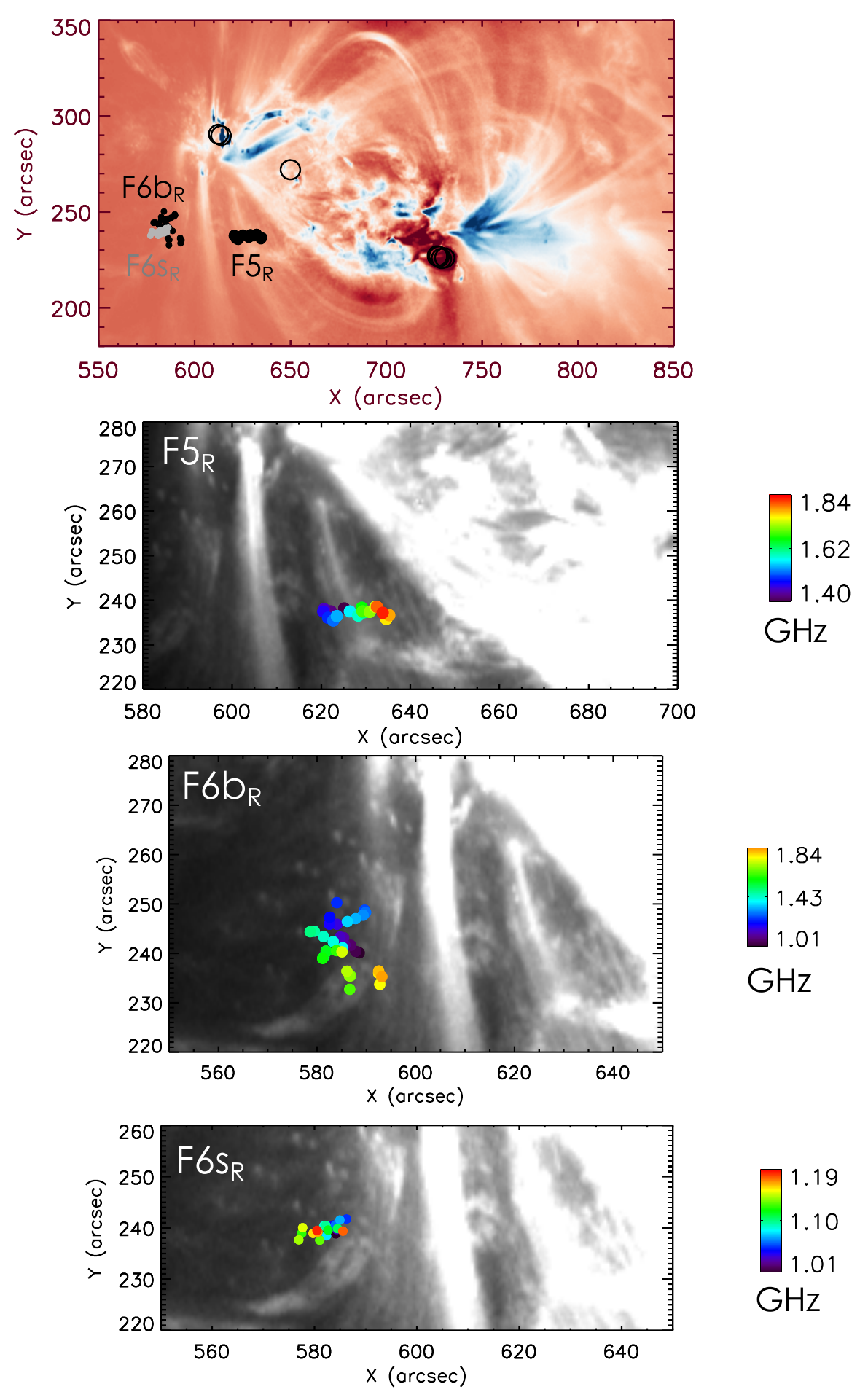}
\caption{Radio source locations as a function of frequency for F5$_R$, F6$_R$ and F6s$_R$. The top image is a composite of a time-averaged AIA 171 \AA\ (red) image and an \ion{Fe}{18} image (blue). Below are close-ups of the centroid locations of three features as a function of frequency overlaid on a AIA171 \AA\ image closest in time to the burst time. The locations as function of frequency are color-coded.}
\label{fig:mag}
\end{figure}

For the brightest radio features (F5$_R$, F6b$_R$, and F6s$_R$) we also investigate the locations as a function of frequency. These are shown in Figure~\ref{fig:mag}, overlaid on a composite AIA image and individually over AIA 171 \AA\  images. The color table in the EUV images was adapted such that faint loops become more visible.  For F5$_R$ and F6b$_R$ we restricted us to centroid locations for which the maximum flux of the source was larger than 0.3 MK. As mentioned before, none of these sources coincide with any of the X-ray sources. F5$_R$ shows a systematic displacement of the centroid location as a function of frequency \textit{perpendicular} to a large, faint loop visible at 171 \AA. F6b$_R$ and F6s$_R$ originate from about the same location, close to a fan-like structure. The scatter of the frequency-dependent centroids in F6b$_R$ is considerable and there is no clear trend, other than that lower frequencies seem to originate preferentially further north than higher frequencies. Within F6s$_R$, the scatter is smaller than in F6b$_R$, but again no systematic displacement of the locations as function of frequency can be observed. 
\subsection{Spectra}
\subsubsection{X-ray spectra}
Table~\ref{t2} summarizes the results from the thermal spectral fitting as described in Section \ref{subs:xrayspec}. As an example of a fitted spectrum, Figure~\ref{fig:bgburst} shows the background-subtracted count rate spectrum of F4. The background-subtracted count rate above 10 keV was around the background level, yet two sources could be imaged in this energy range. The temperatures of the flares ranged between 10.0 MK in F2 to 14.3 MK in F4 and F5. The emission measure was lowest in F7 at $1.2\times 10^{46}$ cm$^{-3}$ and highest in F2 at $4.2\times 10^{46}$ cm$^{-3}$. Even though a nonthermal component could be fitted in F4 and F7, the uncertainties were too large (same order or larger as the fit parameters) for the result to be interpreted with confidence.

\subsubsection{Radiospectra}
The radio spectra shown in Figure \ref{fig:bgburst} show the maximum brightness temperature as a function of frequency in LCP polarization. 
Except for F5$_R$, F6s$_R$, and F6b$_R$ the brightness temperature was low (less than $\sim$ 0.5 MK) or, depending on the frequency, at the noise level of the image. Emission during F5$_R$ was detected at almost all frequencies and peaked at $\sim$ 1.4 GHz. F6b$_R$ had the highest brightness temperature among the broad-band features with a maximum around 2.1 MK at 1 GHz, declining to 0.3 MK at 2 GHz. The maximum brightness temperature of F6s$_R$ (spikes) was 5 MK. Since such bursts may originate from fairly compact source regions, it is likely that they are not spatially resolved. Indeed, the FWHM of the fitted 2D Gaussian was around 20 arcsec at 1.1 GHz while the restoring beam size was 60 arcsec. Hence these values should be considered as lower limit.

\section{Discussion}\label{s:discuss}
During a fifty minute time window, seven X-ray flares between GOES A3.4 and B1.6 (background subtracted) were observed co-temporally with a variety of radio emissions, such as broad-band bursts, narrow-band, short-lived spikes and broad-band continuum emission at frequencies between 1 to 2 GHz. The broad-band continuum emission can be attributed to background thermal emission of the active region, while the short-lived bursts and spikes are indicative of accelerated electrons. The X-ray flares displayed quite typical characteristics regarding size, spatial origin, and energies, as will be discussed in some more detail in Section~\ref{s:xraydisc}. The radio sources were displaced from the X-ray sources by more than 100 arcseconds despite close temporal association with the X-ray emission.  In Section~\ref{s:secondacc} we will discuss possible explanations for this displacement. 
\subsection{X-ray microflare locations and energies}\label{s:xraydisc}
The observed microflares were quite ordinary in the sense that they were short and compact. It is noteworthy that, during F1 and F4, two sources from different locations within the active region were imaged. Since the RHESSI imaging technique favors the brightest source present, this means both of these sources were of comparable brighthness and the X-ray lightcurves and spectra consist of a superposition of emission from two separate flares that were, however, too faint to be analysed individually through imaging spectroscopy. \\
X-ray spectra indicate emission up to $\sim$ 14~keV. For all flares it was possible to fit a thermal component. From the fitted temperature $T$ and emission measure $EM$ one can calculate the thermal energy as 
\begin{equation}
    E_{th}=3k_B T\sqrt{EM\times V}
\end{equation}
where $k_B$ is the Boltzmann constant and $V$ the flaring volume. Assuming a filling factor of unity and a flaring volume of $V=A^{3/2}$ where $A$ is the area of a circle of radius 10 arcsec, one finds thermal energies of the order of $2.4\times 10^{28}$~erg to $4\times 10^{28}$~erg. 
Emission above $\sim$10~keV was at the background level and any spectral fitting model was not well constrained, as described in Section~\ref{xrayspec}. Using upper limit spectral parameters (total electron flux $F_e=1.0\times 10^{34}$s$^{-1}$, electron spectral index $\delta=4.9$ and low-energy cutoff $E_{low}=6.5~$keV) the total nonthermal power is calculated as 
\begin{equation}
    P=\frac{\delta-1}{\delta-2}F_e E_{low}=1.4\times 10^{26} \mathrm{erg \ s^{-1}.}
\end{equation} 
Both, thermal and nonthermal energies are consistent with earlier findings using RHESSI observations for similarly sized flares \citep[][]{2007SoPh..246..339S,2008ApJ...677..704H,2014ApJ...789..116I}.
\subsection{Radio sources as secondary acceleration sites?}\label{s:secondacc}
The most noteworthy observation is the significant displacement between the location of the radio burst sources and the X-ray sources. 
In a simple scenario in which electrons are accelerated during a flare and the signatures of the same accelerated electron population are observed at both, X-ray and radio wavelengths, one would expect the locations of the radio sources to be close to the flaring site as seen in X-rays where the radio emission can be either gyrosynchrotron emission from electrons trapped in flaring loops or coherent emission from electron beams escaping along open field lines visible as e.g. type III bursts. However, only some of the observations presented here can be explained with this simplest scenario. 

Observationally one can distinguish three cases:\\ 
1) Both, X-ray and radio sources, originate from the same magnetic loop structure (F7). \\
2) Radio emission does not originate from the same magnetic structure as X-ray emission, but both structures are magnetically connected (F4, F5).\\
3) Radio emission originates from a different magnetic structure seemingly not connected to the original X-ray flare site (F6). 

\begin{figure*}
\centering
\includegraphics[width=\linewidth]{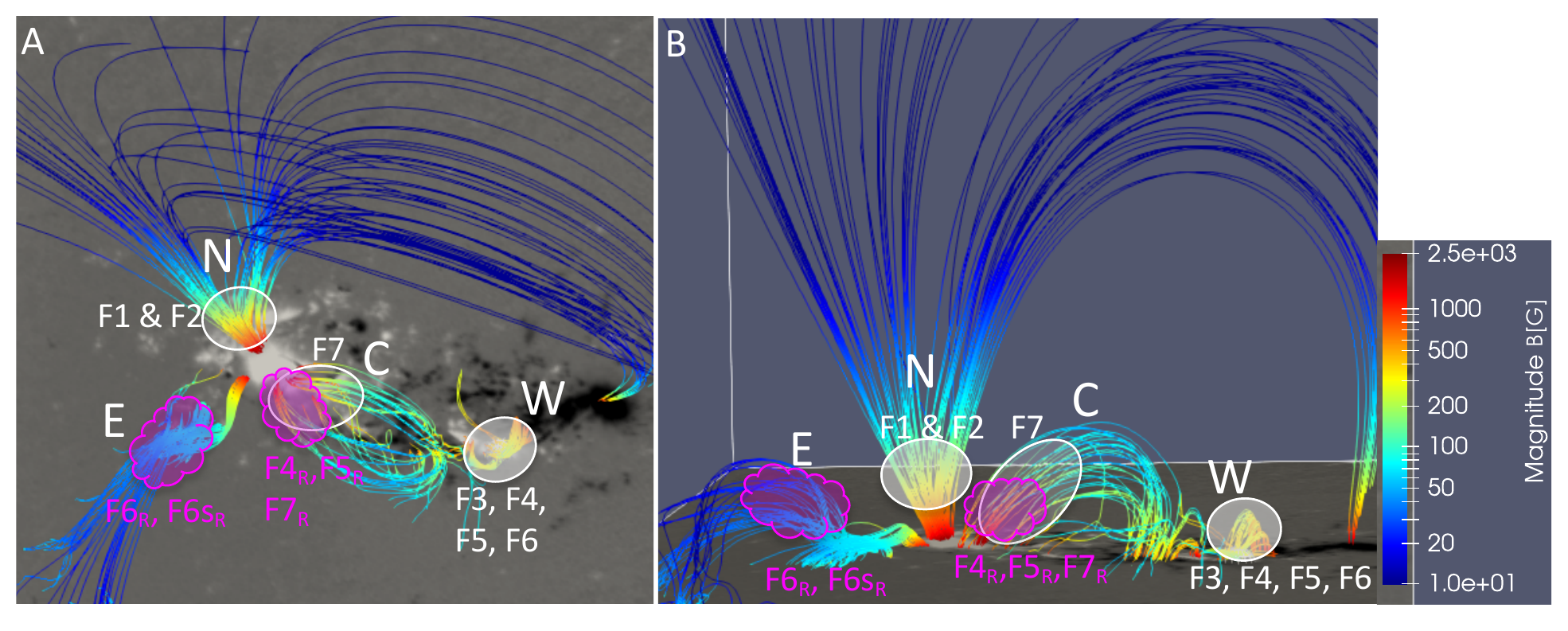}
\caption{Non-linear force free magnetic field extrapolations seen from above (Panel A) and from the side (Panel B). The approximate locations of the X-ray flares are indicated with white ellipses and white labels. The locations of the brightest corresponding radio sources are indicated with purple shapes and labels. }
\label{fig:nlff_mag}
\end{figure*}

To investigate the magnetic connectivity of the active region and confirm the cases outlined above, we performed non-linear force free magnetic field extrapolations (NLFFF) using the {\tt gx\_simulator} package (part of the IDL {\tt SolarSoftware} distribution; \citealt{Nita2015}), shown in Figure~\ref{fig:nlff_mag}.
The extrapolations show four major structures: small, closed magnetic loops at the western side of the active region (labelled W in Figure~\ref{fig:nlff_mag}); large, open and closed field lines in the North (N); a closed loop structure centered between the two mentioned before (C), and another, separate structure to the East (E). Except for the last one, the magnetic loops found by the NLFFF are also visible at EUV wavelengths.
Due to projection effects it is difficult to locate the sources on the magnetic field model exactly and the available data do not provide enough constraints to locate them based on density or height arguments, hence, the likely locations are indicated with elliptic shapes in the Figure, where white ellipses outline the most likely origin of the X-ray emission and purple shapes indicate the likely origin of the corresponding radio emission deduced by qualitative comparison of the EUV features and radio and X-ray source locations. In the following we provide some physical scenarios and explanations for the cases listed above.


\subsubsection{Scenario 1: Same electron population within same magnetic loop (F7)}
During F7, radio emission was observed co-temporally with X-ray emission and, as can be seen in Figures \ref{fig:mag} and \ref{fig:nlff_mag}, the sources at the two different wavelengths originated from the same magnetic loop structure. These observations are compatible with a scenario in which electrons are accelerated during a flare, become trapped in a closed magnetic loop and emit radiation at X-ray and radio wavelengths. 
\subsubsection{Scenario 2: Same electron accelerated electron population with significant transport (F4, F5)}
The radio emission in both of these events was observed co-temporally with the high-energy X-ray emission with the strong burst in F5$_R$ happening 3 seconds after the X-ray peak. The frequency dependent locations of F5$_R$ show no clear trend, indicating that the electrons do not propagate away from the Sun along open field lines as, in that case, one would expect a clear frequency dependence of the source location \citep[e.g.,][]{2013ApJ...763L..21C,2018ApJ...866...62C}. For both of these cases it is likely that the emission is due to electrons that were accelerated near location W and were either directly injected into one of the field lines of loop system C or gained access to it through cross-field diffusion.
\subsubsection{Scenario 3: Secondary electron acceleration (F6b, F6s)}
Both, F6b$_R$ and F6s$_R$ happened between 50 seconds and 54 seconds after the X-ray peak time, and originated from field lines that seemingly are not connected to the X-ray flare site. Short-lived spikes, similar to F6s$_R$, at MHz to GHz frequencies have commonly been interpreted as signatures of flare accelerated electrons due to their high temporal association with hard X-rays \citep[e.g.][]{1986SoPh..104...99B,1991A&A...251..285G,1992ApJ...401..736A}. However, imaging studies of spikes showed that they can be displaced as much as 400 arcsec from the main flare site \citep{2002A&A...383..678B,2006A&A...457..319K, 2009A&A...499L..33B}. Such findings suggested that spikes are signatures of secondary acceleration, that typically happens higher in the Corona but is causally linked to the main energy release. 
However, the connection to the main energy release is not always obvious. Recently, \citet{2021ApJ...911....4L} observed stochastic spike burst associated with an M-class flare for which the connection with the flare was not obvious as is the case for F6b$_R$ and F6s$_R$ in the observations presented here.

Since the NLFFF extrapolated field may deviate from the actual magnetic connectivity in the Corona, particularly during flaring times, one explanation is that we do not capture the fieldlines that connect regions W and E and electrons were accelerated during the flare in region W and transported to region E, similar to the second scenario. However, a more intriguing explanation is that electrons were accelerated in-situ at a secondary acceleration site, triggered by the original flare. Such sympathetic flares, but also coronal mass ejections, filament eruptions and solar energetic electron events have been known for quite some time \citep[e.g.][]{1990A&A...228..513P,2018ApJ...869..177W,2021ApJ...913...89W}, and have been found to even happen between active regions. The exact mechanism that causes them is still under debate. It has been suggested that the regions involved could be connected by large scale coronal loops \citep[e.g.][]{Wang_2001,2008ApJ...677..699J} that are perturbed but the original flare which results in reconnection events far away from the original flare site.


\section{Conclusions}\label{s:conc}
The combined radio and X-ray observations shown here demonstrate that even smaller flares can display a surprising level of complexity. Despite close temporal association between radio and X-ray emission, the two types of emissions originated from entirely different regions. While for some events it can be argued, based on magnetic connectivity, that electrons were accelerated during the primary energy release, in the other events, the radio emissions are likely due to secondary energy release and particle acceleration, either completely unrelated to the primary energy release or triggered by it in a way yet to be understood. Thanks to the high sensitivity of the VLA it is shown that nonthermal processes happen frequently and at many different locations of an active region. Observations from future X-ray instruments that use focusing optics, resulting in a much higher sensitivity and dynamic range as current instruments, combined with VLA observations are needed to fully understand these processes.  

\begin{acknowledgments}

This work was supported by the Swiss National Science Foundation (grant no. 200021\_175832).
It makes use of public archival VLA data from the observing program VLA/13A-384 carried out by the National Radio Astronomy Observatory (NRAO). The authors acknowledge Tim Bastian, Dale Gary, and Stephen White for their help in carrying out the observing program and Gregory Fleishman for help with the GX-simulator package. 
Y.L. and B.C. are supported by NSF grant AGS-1654382 to the New Jersey Institute of Technology.
The NRAO is a facility of the National Science Foundation (NSF) operated under cooperative agreement by Associated Universities, Inc. Some of the figures within this paper were produced using IDL colour-blind-friendly colour tables \citep[see][]{2017zndo....840393W}.
\end{acknowledgments}
\bibliography{mybib}{}

\end{document}